\RequirePackage{fix-cm}
\documentclass[smallextended]{svjour3}       
\usepackage{graphicx}
\usepackage{subcaption}
\usepackage{amssymb}
\usepackage{amsmath}
\usepackage{amsfonts}
\usepackage{mathtools}
\usepackage{multirow}
\usepackage{bm}
\usepackage{nicefrac}
\usepackage{tabularx}
\usepackage{booktabs}
\usepackage{pifont}
\usepackage{extarrows}
\usepackage{balance}
\usepackage[ruled,vlined]{algorithm2e}
\usepackage{array}

\makeatletter
\def\hlinewd#1{%
  \noalign{\ifnum0=`}\fi\hrule \@height #1 \futurelet
   \reserved@a\@xhline}
\makeatother

\SetKwInput{KwInput}{Input}                
\SetKwInput{KwOutput}{Output}              

\newcommand{\sudo}{SuDoRM-RF }
\newcommand{\sudoi}{SuDoRM-RF++ }
\newcommand{\csudoi}{C-SuDoRM-RF++ }
\newcommand{\sudodot}{SuDoRM-RF}
\newcommand{\sudodoti}{SuDoRM-RF++}
\newcommand{\csudodoti}{C-SuDoRM-RF++}

\newcommand{\sudoxl}{SuDoRM-RF 2.0x }
\newcommand{\sudol}{SuDoRM-RF 1.0x }
\newcommand{\sudom}{SuDoRM-RF 0.5x }
\newcommand{\sudos}{SuDoRM-RF 0.25x }
\newcommand{\sudoixl}{SuDoRM-RF++ 2.0x }
\newcommand{\sudoil}{SuDoRM-RF++ 1.0x }

\newcommand{\csudoim}{C-SuDoRM-RF++ 0.5x }
\newcommand{\csudois}{C-SuDoRM-RF++ 0.25x }

\newcommand\srelu{\operatorname{ReLU}}

\newcommand\lp{\left(}
\newcommand\rp{\right)}

\newcommand\y{\mathbf{y}}
\newcommand\x{\mathbf{x}}
\newcommand\T{T}
\newcommand\CE{C_{\mathcal{E}}}
\newcommand\Cout{C}
\newcommand\Cin{C_U}
\newcommand\Kin{K_U}
\newcommand\Sin{S_U}

\newcommand\KE{K_{\mathcal{E}}}

\newcommand\R{\mathbb{R}}
\newcommand\vx{\mathbf{v}_{\mathbf{x}}}
\newcommand\evi{\widehat{\mathbf{v}}_{i}}
\newcommand\emi{\widehat{\mathbf{m}}_{i}}
\newcommand\esi{\widehat{\mathbf{s}}_{i}}

\newcommand\s{\mathbf{s}}

\begin{document}

\title{Compute and memory efficient universal sound source separation}


\author{Efthymios Tzinis \and Zhepei Wang \and \\ Xilin Jiang \and Paris Smaragdis 
}


\institute{Efthymios Tzinis \at
             University of Illinois at Urbana-Champaign \\
            \email{etzinis2@illinois.edu}           
          \and
          Zhepei Wang \at University of Illinois at Urbana-Champaign \and
          Xilin Jiang \at University of Illinois at Urbana-Champaign \and
          Paris Smaragdis \at University of Illinois at Urbana-Champaign \& Adobe Research
}

\date{Received: date / Accepted: date}

\maketitle

\begin{abstract}
Recent progress in audio source separation led by deep learning has enabled many neural network models to provide robust solutions to this fundamental estimation problem. In this study, we provide a family of efficient neural network architectures for general purpose audio source separation while focusing on multiple computational aspects that hinder the application of neural networks in real-world scenarios. The backbone structure of this convolutional network is the SUccessive DOwnsampling and Resampling of Multi-Resolution Features (\sudodot) as well as their aggregation which is performed through simple one-dimensional convolutions. This mechanism enables our models to obtain high fidelity signal separation in a wide variety of settings where a variable number of sources are present and with limited computational resources (e.g. floating point operations, memory footprint, number of parameters and latency). Our experiments show that \sudo models perform comparably and even surpass several state-of-the-art benchmarks with significantly higher computational resource requirements. The causal variation of \sudo is able to obtain competitive performance in real-time speech separation of around $10$dB scale-invariant signal-to-distortion ratio improvement (SI-SDRi) while remaining up to $20$ times faster than real-time on a laptop device. 
\end{abstract}

\section{Introduction}
\label{sec:intro}

The advent of the deep learning era has enabled the effective usage of neural networks towards single-channel source separation with mask-based architectures \cite{huang2014deep}. Recently, end-to-end source separation in time-domain has shown state-of-the-art results in a variety of separation tasks such as speech separation \cite{luo2019convTasNet,luo2019dual}, universal sound separation \cite{kavalerov2019universal,tzinis2019improving} and music source separation \cite{defossez2019demucs}. The separation module of ConvTasNet \cite{luo2019convTasNet} and its variants \cite{kavalerov2019universal,tzinis2019improving} consist of multiple stacked layers of depth-wise separable convolutions \cite{sifre2014depthwiseseparable} which can aptly incorporate long-term temporal relationships. Building upon the effectiveness of a large temporal receptive field, a dual-path recurrent neural network (DPRNN) \cite{luo2019dual} has shown remarkable performance on speech separation. Demucs \cite{defossez2019demucs} has a refined U-Net structure \cite{ronneberger2015original_unet} and has shown strong performance improvement on music source separation. Specifically, it consists of several convolutional layers in each a downsampling operation is performed in order to extract high dimensional features. A two-step approach has been introduced in \cite{tzinis2019two} and showed that universal sound separation models could be further improved when working directly on the latent space and learning the ideal masks on a separate step.

Despite the dramatic advances in source separation performance, the computational complexity of the aforementioned methods might hinder their extensive usage across multiple devices. Specifically, many of these algorithms are not amenable to, e.g., embedded systems deployment, or other environments where computational resources are constrained.  Additionally, training such systems is also an expensive computational undertaking which can amount to significant costs.

Several studies, mainly in the image domain, have introduced more efficient architectures in order to overcome the growing concern of large models with high computational requirements. Models with depth-wise separable convolutions \cite{sifre2014depthwiseseparable} have shown strong potential for several image-domain tasks \cite{chollet2017xception_depthwiseseparable} while significantly reducing the computational requirements. Thus, several variants such as MobileNets \cite{howard2017mobilenets} have been proposed for deep learning on edge devices. However, convolutions with a large dilation factor might inject several artifacts and thus, lightweight architectures that combine several dilation factors in each block have been proposed for image tasks \cite{mehta2019espnetv2}. More recent studies propose meta-learning algorithms for optimizing architecture configurations given specific computational resource and accuracy requirements \cite{yu2019slimmablenets,cai2019onceandforall}.

Despite the recent success of low-resource architectures in the image domain, little progress has been made towards proposing efficient architectures for audio tasks and especially source separation. In \cite{kalchbrenner2018efficient_audiosynthesis} a WaveRNN is used for efficient audio synthesis in terms of floating point operations (FLOPs) and latency. Other studies have introduced audio source separation models with reduced number of trainable parameters \cite{luo2019dual,maldonado2020lightweight,luo2020groupcomm} and binarized models \cite{kim2018bitwise}. Modern approaches, mainly in speech enhancement and music source separation, have been focusing on developing models which are capable of real-time inference. Specifically, a temporal convolutional network for real time speech enhancement has been proposed in \cite{pandey2019tcnnRealTimeSpeechEnhancement} while the latest state-of-the-art performance has been obtained by a real-time variation of Demucs for online speech denoising \cite{defossez2020realtimedemucs}. In a similar sense real-time music source separation models have been proposed in \cite{hennequin2020spleeterRealTimeMusicSep} and a system capable of real-time speech separation from background music has been implemented in \cite{kaspersen2020hydranetRealTimeSpeechFromMusic}.

In this study, we propose a novel efficient neural network architecture for audio source separation while following a more holistic approach in terms of computational resources that we take into consideration (FLOPs, latency and total memory requirements). Our proposed model performs SUccessive DOwnsampling and Resampling of Multi-Resolution Features (\sudodot) using depth-wise convolutions. By doing so, \sudo exploits the effectiveness of iterative temporal resampling strategies \cite{haris2018deepbackprojectionSuperResolution} and avoids the need for multiple stacked dilated convolutional layers \cite{luo2019convTasNet}. We also propose improved versions of the aforementioned architecture with significant benefits in terms of computational resource requirements as well as causal variations where online inference is available. We report a separation performance comparable to or even better than several recent state-of-the-art models on speech, environmental and universal sound separation tasks with significantly lower computational requirements. Our experiments suggest that \sudo models a) could be deployed on devices with limited resources, b) be trained significantly faster and achieve good separation performance and c) scale well when increasing the number of parameters. Our code is available online \footnote{Code: https://github.com/etzinis/sudo\_rm\_rf}.

\begin{figure}[!htb]
    \centering
  \begin{subfigure}[h]{\linewidth}
      \includegraphics[width=\linewidth]{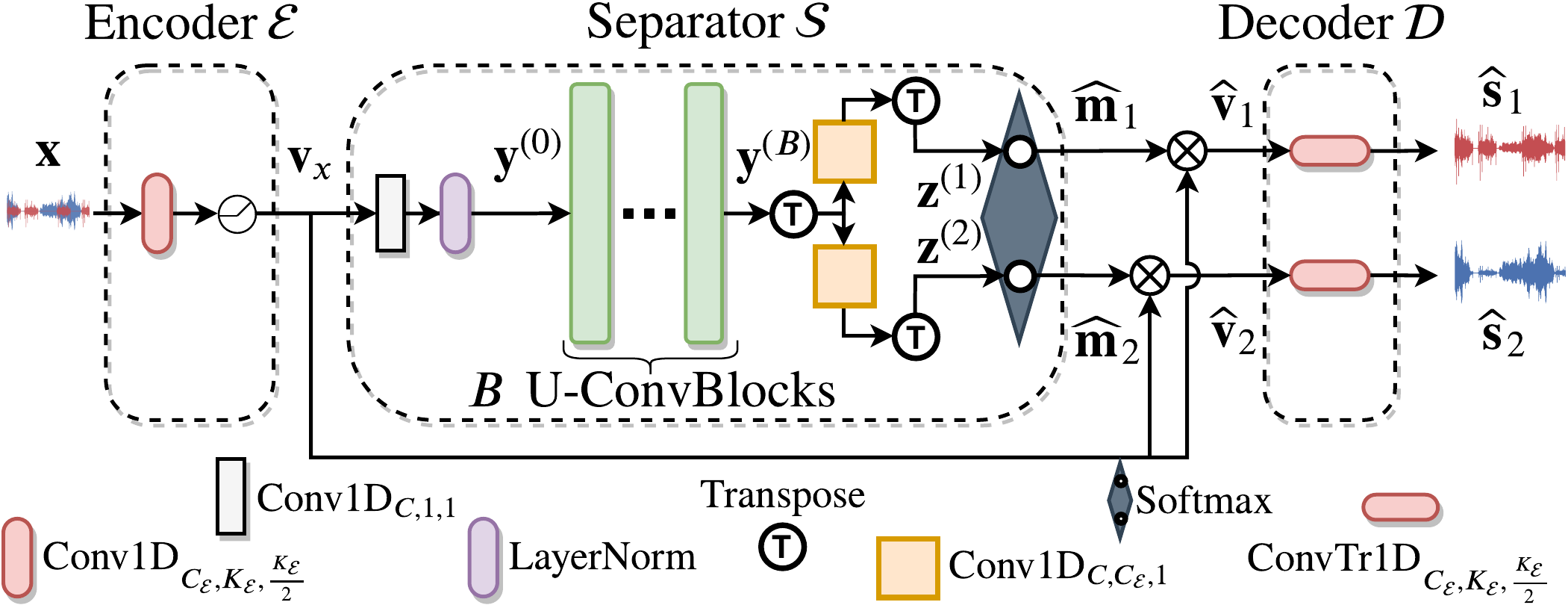}
      \caption{\sudo architecture.}
      \label{fig:sudormrf} 
     \end{subfigure} \\
 \begin{subfigure}[h]{\linewidth}
      \includegraphics[trim={2.5cm 0 6cm 0},width=\linewidth]{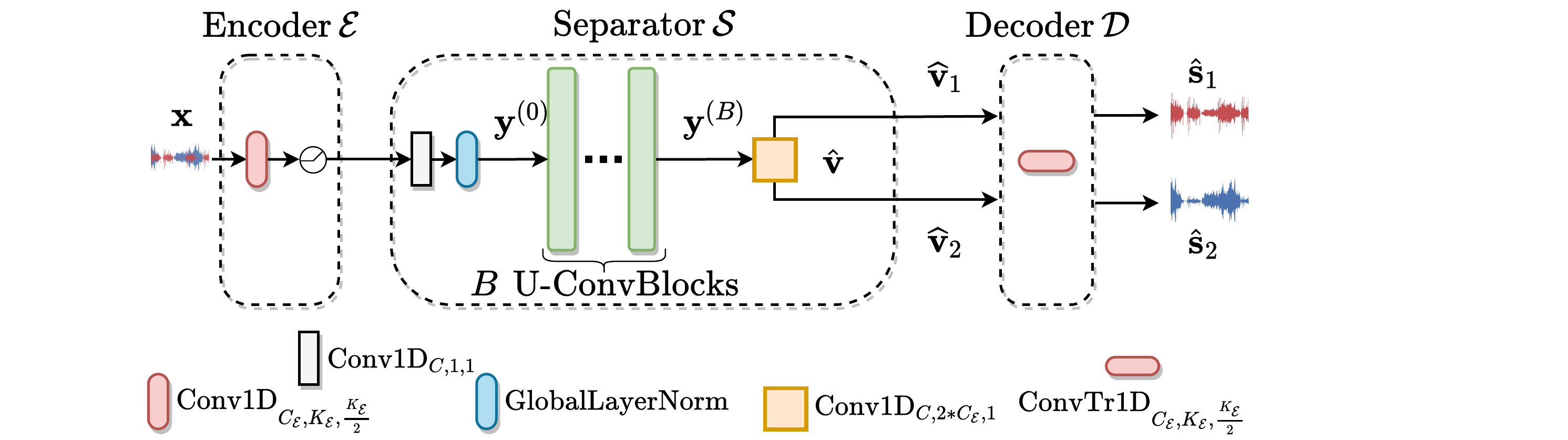}
      \caption{\sudoi architecture.}
      \label{fig:sudormrfi} 
     \end{subfigure} \\
    \caption{\sudo and \sudoi architectures for separating two sources.}
    \label{fig:sudormrf_all}
\end{figure} 

\section{Sudo rm -rf network architecture}
\label{sec:net_arch}
On par with many state-of-the-art approaches in the literature \cite{luo2019convTasNet,tzinis2019two,luo2019dual,defossez2019demucs}, \sudo performs end-to-end audio source separation using a mask-based architecture with adaptive encoder and decoder basis. We have extended our basic model in order to also remove the mask estimation process by introducing \sudoi that directly estimates the latent representations of the sources in the adaptive front-end domain. First, we describe all the modules which are needed for both architectures and describe extensively the inference path for our basic \sudo architecture. In Figure \ref{fig:sudormrf_all}, both architectures are shown. Consequently we also present the extensions of our original \sudo model including: its improved version \sudoi (Section \ref{sec:net_arch:sudoi}), a variant of \sudoi including group communication \cite{luo2020groupcomm} (Section \ref{sec:net_arch:sudoigc}) as well as its causal variation \csudoi (Section \ref{sec:net_arch:csudoi}). 

The input is the raw signal from a mixture $\textbf{x} \in \R^T$ with $T$ samples in the time-domain. First, we feed the input mixture $\textbf{x}$ to an encoder $\mathcal{E}$ in order to obtain a latent representation for the mixture $\vx = \mathcal{E} \left( \textbf{x} \right) \in \R^{\CE \times L}$. Consequently the latent mixture representation is fed through a separation module $\mathcal{S}$ which estimates the corresponding masks $\emi \in \R^{\CE \times L}$ for each one of the $N$ sources $\s_1, \cdots, \s_N \in \R^T$ which constitute in the mixture. The estimated latent representation for each source in the latent space $\evi$ is retrieved by multiplying element-wise an estimated mask $\emi$ with the encoded mixture representation $\vx$. Finally, the reconstruction for each source $\esi$ is obtained by using a decoder $\mathcal{D}$ to transform the latent-space $\evi$ source estimates back into the time-domain $\esi = \mathcal{D} \left( \evi \right)$. An overview of the \sudo architecture is displayed in Figure \ref{fig:sudormrf}. The encoder, separator and decoder modules are described in Sections \ref{sec:net_arch:encoder}, \ref{sec:net_arch:separator} and \ref{sec:net_arch:decoder}, respectively. For simplicity of our notation we will describe the whole architecture assuming that the processed batch size is one. Moreover, we are going to define some useful operators of the various convolutions which are used in \sudodot.

    

\begin{definition}
$\operatorname{Conv1D}_{C, K, S}:\R^{C_{in} \times L_{in}} \rightarrow \R^{C \times L}$ defines a kernel $\mathbf{W} \in \R^{C \times C_{in} \times K}$ and a bias vector $\mathbf{b} \in \R^{C}$. When applied on a given input $\mathbf{x} \in \R^{C_{in} \times L_{in}}$ it performs a one-dimensional convolution operation with stride equal to $S$ as shown next:
\begin{equation}
\label{eq:conv1d}
    \begin{gathered}
    \operatorname{Conv1D}_{C, K, S}\lp \x \rp_{i,l} = \mathbf{b}_i + \sum_{j=1}^{C_{in}} \sum_{k=1}^{K} \mathbf{W}_{i,j,k}\cdot \x_{j, S\cdot l- k},
    \end{gathered}
\end{equation}
where the indices $i$, $j$, $k$, $l$ denotes the output channel, the input channel, the kernel sample and, the temporal index, respectively.  Note that without loss of generality and performing appropriate padding, the last dimension of the output representation would be $L = \left \lfloor{\nicefrac{L_{in}}{S}}\right \rfloor$. 
\end{definition}

\begin{definition}
$\operatorname{ConvTr1D}_{C, K, S}:\R^{C_{in} \times L_{in}} \rightarrow \R^{C \times L}$ defines a one-dimensional transpose convolution. Since any convolution operation could be expressed as a matrix multiplication, transposed convolution can be directly understood as the gradient calculation for a regular convolution w.r.t. its input \cite{Simonyan2013DeepIC_transposeconvolution}.
\end{definition}

\begin{definition}
$\operatorname{DWConv1D}_{C, K, S}:\R^{C_{in} \times L_{in}} \rightarrow \R^{C \times L}$ defines a one-dimensional depth-wise convolution operation \cite{sifre2014depthwiseseparable}. In essence, this operator defines $G=C_{in}$ separate one-dimensional convolutions $\mathcal{F}_i = \left[ \operatorname{Conv1D}_{C_{G}, K, S} \right]_i$ with $i \in \{1, \cdots, G \}$ where $C_{G} = \left \lfloor{\nicefrac{C}{G}}\right \rfloor$. Given an input $\mathbf{x} \in \R^{C_{in} \times L_{in}}$ the $i$th one-dimensional convolution contributes to $C_{G} = \left \lfloor{\nicefrac{C}{G}}\right \rfloor$ output channels by considering as input only the $i$th row of the input as described below: 
\begin{equation}
\label{eq:dwconv1d}
    \begin{gathered}
    \operatorname{DWConv1D}_{C, K, S}\lp \x \rp = \operatorname{Concat}\lp \lbrace \mathcal{F}_i\lp \x_i \rp, \enskip \forall i \rbrace \rp,
    \end{gathered}
\end{equation}
where $\operatorname{Concat}(\cdot)$ performs the concatenation of all individual one-dimensional convolution outputs across the channel dimension. 
\end{definition}

\subsection{Encoder}
\label{sec:net_arch:encoder}
The encoder $\mathcal{E}$ architecture consists of a one-dimensional convolution with kernel size $\KE$ and stride equal to $\nicefrac{\KE}{2}$ similar to \cite{luo2019convTasNet}. Each convolved input audio segment of $\KE$ samples is transformed to a $\CE$-dimensional vector representation where $\CE$ is the number of output channels of the $1D$-convolution. We force the output of the encoder to be strictly non-negative by applying a rectified linear unit (ReLU) activation on top of the output of the $1D$-convolution. Thus, the encoded input mixture representation could be expressed as:
\begin{equation}
\label{eq:encoder}
    \begin{gathered}
    \vx = \mathcal{E} \lp \textbf{x} \rp = \srelu \lp  \operatorname{Conv1D}_{\CE, \KE, \nicefrac{\KE}{2}} \lp \textbf{x}\rp  \rp  \in \R^{\CE \times L},
    \end{gathered}
\end{equation}
where the activation $\srelu \lp \cdot \rp $ is applied element-wise.

\subsection{Separator}
\label{sec:net_arch:separator}
In essence, the separator $\mathcal{S}$ module performs the following transformations to the encoded mixture representation $\vx \in \R^{\CE \times L}$:
\begin{enumerate}
    \item Projects the encoded mixture representation $\vx \in \R^{\CE \times L}$ to a new channel space through a layer-normalization (LN) \cite{ba2016layernormalization} followed by a point-wise convolution as shown next:
    \begin{equation}
    \label{eq:sep0}
    \begin{gathered}
    \y_0 = \operatorname{Conv1D}_{\Cout, 1, 1} \lp \operatorname{LN}\lp \vx \rp  \rp   \in \R^{\Cout \times L},
    \end{gathered}
    \end{equation}
    where $\operatorname{LN}\lp \vx \rp$ denotes a layer-normalization layer in which the moments used are extracted across the temporal dimension for each channel separately.
    \item Performs repetitive non-linear transformations provided by $B$ U-convolutional blocks (U-ConvBlocks) on the intermediate representation $\y_0$. In other words, the output of the $i$th U-ConvBlock would be denoted as $\y_i \in \R^{\Cout \times L}$ and would be used as input for the $(i+1)$th block. Each U-ConvBlock extracts and aggregates information from multiple resolutions which is extensively described in Section \ref{sec:net_arch:separator:main_block}.  
    \item Aggregates the information over multiple channels by applying a regular one-dimensional convolution for each source on the transposed feature representation $\y_B^\T \in \R^{L \times \Cout}$. Effectively, for the $i$th source we obtain an intermediate latent representation as shown next:
    \begin{equation}
    \label{eq:premasks}
    \begin{gathered}
    \mathbf{z}_i = \operatorname{Conv1D}_{\Cout, \CE, 1} \lp  \y_B^\T  \rp^\T   \in \R^{\CE \times L}
    \end{gathered}
    \end{equation}
    
    This step has been introduced in \cite{tzinis2019two} and empirically shown to make the training process more stable rather than using the activations from the final block $\y_B$ to estimate the masks.
    \item Combines the aforementioned latent codes for all sources $\mathbf{z}_i \enskip \forall i \in \{1, \cdots, N\}$ by performing a softmax operation in order to get mask estimates $\emi \in [0,1]^{\CE \times L}$ which add up to one across the dimension of the sources. Namely, the corresponding mask estimate for the $i$th source would be:
    \begin{equation}
    \label{eq:masks}
    \begin{gathered} 
    \emi = \operatorname{vec}^{-1} \lp \frac{\exp \lp \operatorname{ \operatorname{vec} \lp \mathbf{z}_i \rp} \rp } {\sum_{j=1}^{N} \exp \lp \operatorname{vec} \lp \mathbf{z}_j \rp \rp } \rp \in \R^{ \CE \times L},
    \end{gathered}
    \end{equation}
    where $\operatorname{vec} \lp \cdot \rp: \R^{K \times N} \rightarrow \R^{K \cdot N} $ and $\operatorname{vec}^{-1} \lp \cdot \rp: \R^{K \cdot N} \rightarrow \R^{K \times N} $ denotes the vectorization of an input tensor and the inverse operation, respectively. 
    \item Estimates a latent representation $\evi \in \R^{ \CE \times L}$ for each source by multiplying element-wise the encoded mixture representation $\vx$ with the corresponding mask $\emi$:
    \begin{equation}
    \label{eq:estimated_latent}
    \begin{gathered} 
    \evi = \vx \odot \emi \in \R^{ \CE \times L},
    \end{gathered}
    \end{equation}
    where $\mathbf{a} \odot \mathbf{b}$ is the element-wise multiplication of the two tensors $\mathbf{a} $ and $ \mathbf{b}$ assuming that they have the same shape. 
\end{enumerate}
%
\begin{figure}[!htb]
    \centering
      \includegraphics[width=\linewidth]{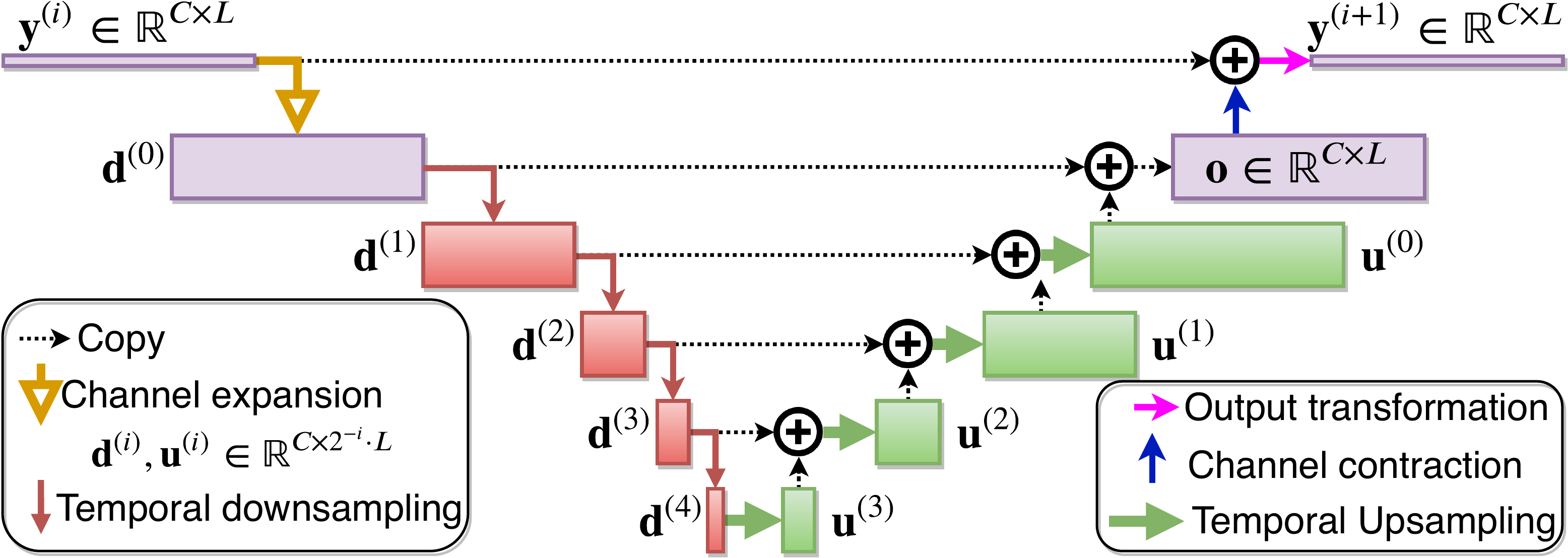}
      \caption{U-ConvBlock architecture.}
      \label{fig:uconvblock}
\end{figure}


\begin{algorithm}[t!]
\SetAlgoLined
\KwInput{$\y^{(i)} \in \R^{\Cout \times L}$} 
\KwOutput{$\y^{(i+1)} \in \R^{\Cout \times L}$} 
\tcp{Expand channel dimensions}
 $\mathbf{q} \gets \operatorname{PReLU}_{\Cin} \lp \operatorname{LN} \lp \operatorname{Conv1D}_{\Cin, 1, 1} \lp \y^{(i)} \rp \rp \rp $\;
 $\mathbf{d}^{(0)} \gets \operatorname{PReLU}_{\Cin} \lp \operatorname{LN} \lp \operatorname{DWConv1D}_{\Cin, \Kin, 1} \lp \mathbf{q} \rp \rp \rp $\;
 \For{$i = 1$; $i{+}{+}$; while $i <= Q$}{
 \tcp{Successive depth-wise downsampling}
  $\mathbf{d}^{(i)} \gets \operatorname{LN} \lp \operatorname{DWConv1D}_{\Cin, \Kin, \Sin} \lp \mathbf{d}^{(i-1)} \rp \rp $\;
  $\mathbf{d}^{(i)} \gets \operatorname{PReLU}_{\Cin} \lp \mathbf{d}^{(i)}  \rp $\;
 }
 $\mathbf{u}^{(Q)} \gets \mathbf{d}^{(Q)}$\;
 \For{$i = Q - 1$; $i{-}{-}$; while $i >= 0$}{
 \tcp{Upsampling and adding multi-resolution features}
  $\mathbf{u}^{(i)} \gets \mathbf{d}^{(i)} + \mathcal{I}_{\Sin} \lp \mathbf{u}^{\lp i+1\rp} \rp $\;
 }
 $\mathbf{o} \gets  \operatorname{LN} \lp \operatorname{Conv1D}_{\Cout, 1, 1} \lp \operatorname{PReLU}_{\Cout} \lp \operatorname{LN} \lp \mathbf{u}^{(0)} \rp \rp \rp \rp $\;
 \Return $ \operatorname{PReLU}_{\Cout} \lp \y^{(i)} + \mathbf{o}  \rp$\;
 \caption{U-ConvBlock forward pass}
 \label{alg:uconvblock}
\end{algorithm}

\subsubsection{U-convolutional block (U-ConvBlock)}
\label{sec:net_arch:separator:main_block}
U-ConvBlock uses a block structure which resembles a depth-wise separable convolution \cite{sifre2014depthwiseseparable} with a skip connection as in ConvTasNet \cite{luo2019convTasNet}. However, instead of performing a regular depth-wise convolution as shown in \cite{chollet2017xception_depthwiseseparable} or a dilated depth-wise which has been successfully utilized for source separation \cite{luo2019convTasNet,tzinis2019improving,tzinis2019two} our proposed U-ConvBlock extracts information from multiple resolutions using $Q$ successive temporal downsampling and $Q$ upsampling operations similar to a U-Net architecture \cite{ronneberger2015original_unet}. More importantly, the output of each block leaves the temporal resolution intact while increasing the effective receptive field of the network multiplicatively with each temporal sub-sampling operation \cite{luo2016effectivereceptivefield}. We postulate that this whole resampling procedure of extracting features at multiple scales combined with the efficient increase of the effective receptive field enables \sudo models to outperform several convolutional architectures and perform in par with much more expensive recurrent and self-attention architectures \cite{subakan2020attentionSeparation}. An abstract view of the $i$th U-ConvBlock is displayed in Figure \ref{fig:uconvblock} while a detailed description of the operations is presented in Algorithm \ref{alg:uconvblock}. 
\begin{definition}
$\operatorname{PReLU}_{C}:\R^{C \times L} \rightarrow \R^{C \times L}$ defines a parametric rectified linear unit (PReLU) \cite{he2015PReLU} with $C$ learnable parameters $\mathbf{a}  \in  \R^{C} $. When applied to an input matrix $\y \in \R^{C \times L}$ the non-linear transformation could be defined element-wise as:
\begin{equation}
\label{eq:prelu}
    \begin{gathered}
    \operatorname{PReLU}_{C}\lp \y \rp_{i,j} = \operatorname{max} \lp 0, \y_{i,j}  \rp + \mathbf{a}_i \cdot \operatorname{min} \lp 0, \y_{i,j}  \rp 
    \end{gathered}
\end{equation}
\end{definition}
\begin{definition}
$\mathcal{I}_{M}:\R^{C \times L} \rightarrow \R^{C \times M \cdot L}$ defines a nearest neighbor temporal interpolation by a factor of $M$. When applied on an input matrix $\y \in \R^{C \times L}$ this upsampling procedure could be formally expressed element-wise as: $\mathcal{I}_{M}\lp \mathbf{u} \rp_{i,j} = \mathbf{u}_{i, \left \lfloor{\nicefrac{j}{M}}\right \rceil}$
\end{definition}

\begin{definition}
$\operatorname{LN}:\R^{C \times L} \rightarrow \R^{C \times L}$ defines a parametric normalization layer \cite{ba2016layernormalization} with learnable parameters $\mathbf{\gamma} \in  \R^{C} $ and $\mathbf{\beta} \in  \R^{C} $. When applied to an input matrix $\y \in \R^{C \times L}$ the normalization could be defined element-wise as:
\begin{equation}
\label{eq:ln}
    \begin{gathered}
    \operatorname{LN} \lp \y \rp_{i,j} = \frac{\y_{i,j} - \mu_i}{\sigma_i} \gamma_i + \beta_i, \enskip 
    \mu_i = \sum_{j} \y_{i,j}, \enskip \sigma_i = \sqrt{\sum_{j} \lp \y_{i,j} - \mu_i \rp^2}  
    \end{gathered}
\end{equation}
\end{definition}

\begin{definition}
$\operatorname{GLN}:\R^{C \times L} \rightarrow \R^{C \times L}$ defines a parametric normalization layer with learnable parameters $\mathbf{\gamma} \in  \R^{C} $ and $\mathbf{\beta} \in  \R^{C} $. When applied to an input matrix $\y \in \R^{C \times L}$ the normalization could be defined element-wise as:
\begin{equation}
\label{eq:gln}
    \begin{gathered}
    \operatorname{GLN} \lp \y \rp_{i,j} = \frac{\y_{i,j} - \mu}{\sigma} \gamma_i + \beta_i, \enskip 
    \mu = \sum_{i,j} \y_{i,j}, \enskip \sigma = \sqrt{\sum_{i,j} \lp \y_{i,j} - \mu \rp^2}  
    \end{gathered}
\end{equation}
\end{definition}

\subsection{Decoder}
\label{sec:net_arch:decoder}
Our decoder module $\mathcal{D}$ is the final step in order to transform the latent space representation $\evi$ for each source back to the time domain. In our proposed model we follow a similar approach as in \cite{tzinis2019two} where each latent source representation $\evi$ is fed through a different transposed convolution decoder $\operatorname{ConvTr1D}_{\CE, \KE, \nicefrac{\KE}{2}}$. The efficacy of dealing with different types of sources using multiple decoders has also been studied in \cite{differentdecoders}. Ignoring the permutation problem, for the $i$th source we have the following reconstruction in time: 
\begin{equation}
\label{eq:decoder}
    \begin{gathered}
    \esi = \mathcal{D}_i \left( \evi \right) = \operatorname{ConvTr1D}_{\CE, \KE, \nicefrac{\KE}{2}} \lp \evi \rp
    \end{gathered}
\end{equation}

\subsection{Improved version with no mask estimation \sudoi}
\label{sec:net_arch:sudoi}
In the improved version of the proposed architecture, namely, \sudoi, the model estimates directly the latent representation for each target signal $\evi \in \R^{ \CE \times L}$ and uses only one decoder module. Our intuition lies in the aspect that a highly parameterized neural network could potentially estimate those targets without the need of the hard-regularized element-wise multiplication process of the masks on top of the mixture encoded representation $\vx \in \R^{ \CE \times L}$. Essentially, \sudodoti, which is presented in Figure \ref{fig:sudormrfi}, can be derived from our initial \sudodot model by applying the following alternations to the architecture:
\begin{itemize}
\item We replace the mask estimation and element-wise multiplication process with a direct estimation of the latent target signals $\evi$ after the final output of the model $\y^{(B)}$. We have validated experimentally that by removing the mask estimation layer leads to similar or slightly improved results.
\item We use only one trainable decoder in order to transform the latent representation back to the time domain instead of two separate ones, namely, $\esi = \mathcal{D} \left( \evi \right) = \operatorname{ConvTr1D}_{\CE, \KE, \nicefrac{\KE}{2}} \lp \evi \rp$.
\item We replace the layer normalization layers (Equation \ref{eq:ln}) with global layer normalization (GLN) layers as defined in (Equation \ref{eq:gln}). This change significantly improves the convergence of our models probably because of the interdependence of the gradient statistics between the channels.
\item For each intermediate representation with $C$ channels, we simplify the activation layers and we use PReLU activation layers with only one learnable parameter instead of $C$ as it was initially defined in Equation \ref{eq:prelu}. In this way, we are able to achieve similar results as before with less parameters.
\end{itemize}
We would like to underline that the structure of the initial \sudo models could potentially outperform the improved \sudoi variation in cases where the direct estimation of the latent targets would be more difficult than estimating the masks (e.g. unconstrained optimization in those latent spaces might be worse than estimating a bounded mask with values in the $[0, 1]$ region). Moreover, the alternation of containing two decoders proposed in the initial version might be more useful in cases where one wants to solve an audio source separation problem containing two distinct classes of sounds (e.g. speech enhancement) where each decoder could be fine-tuned towards decoding the class-specific characteristics of each estimated latent representation.

\subsection{Group communication variation}
\label{sec:net_arch:sudoigc}
We also propose a new variation of our model, namely, \sudoi GC, where we combine group communication (GC) with our improved version of our model \sudodoti. GC is a novel way to significantly reduce the parameters of an audio processing network which has been recently proposed in \cite{luo2020groupcomm}. In the proposed architecture, the intermediate representations are being processed in groups of sub-bands of channels. We divide the channels of each $1 \times 1$ convolutional block into $16$ groups and we process them first independently by sharing the parameters across all groups of sub-bands. At a second step, we apply a self-attention module \cite{shaw2018selfattention} to combine them. The resulting architecture leads to a significant improvement in the number of trainable parameters which is mainly dominated by the bottleneck dense layers.

\subsection{Causal version \csudoi}
\label{sec:net_arch:csudoi}

\begin{figure}[!htb]
    \centering
  \begin{subfigure}[h]{0.49\linewidth}
      \includegraphics[width=\linewidth]{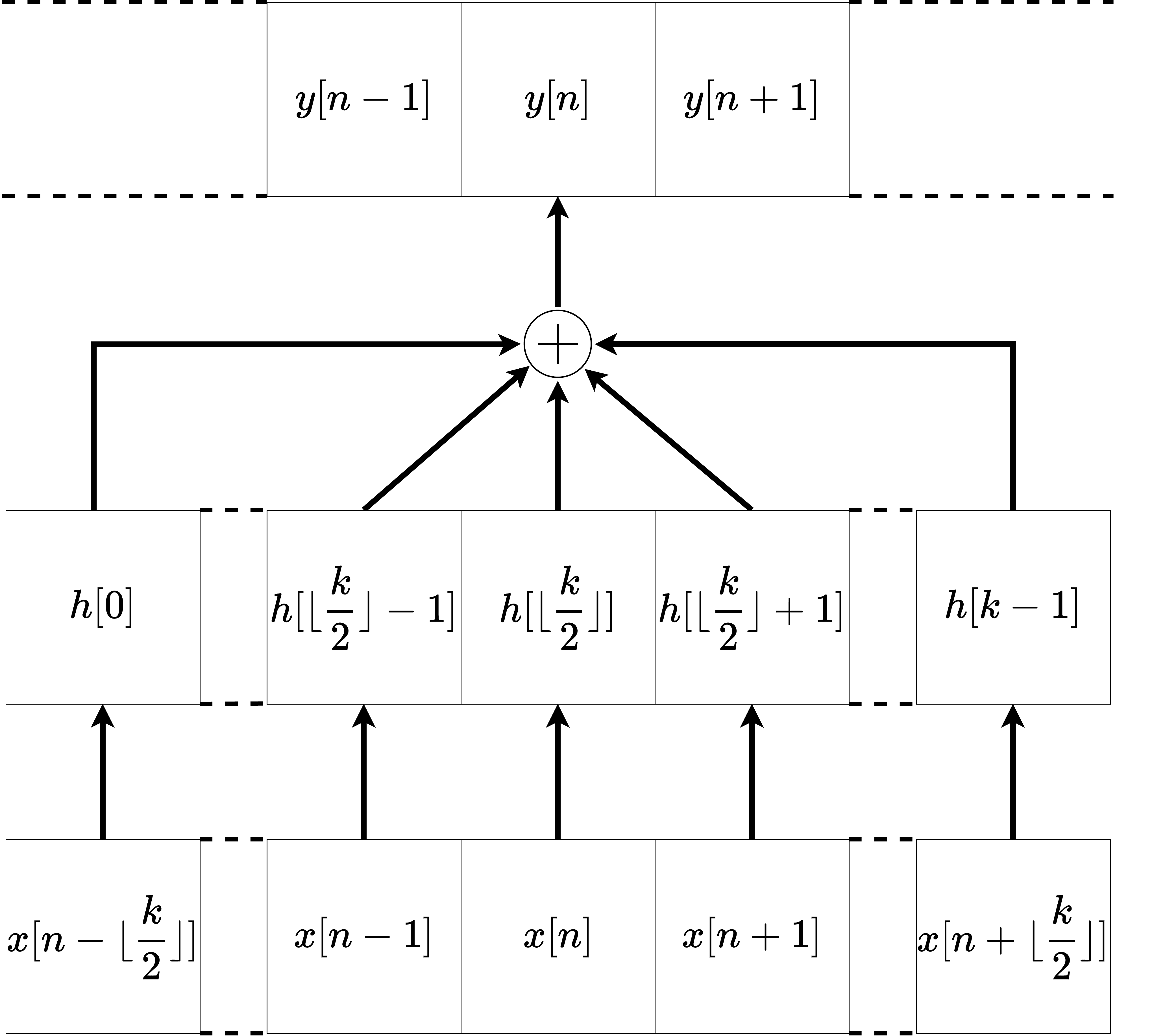}
      \caption{Non-causal convolution.}
      \label{fig:noncausalconv} 
     \end{subfigure}
  \begin{subfigure}[h]{0.47\linewidth}
      \includegraphics[width=\linewidth]{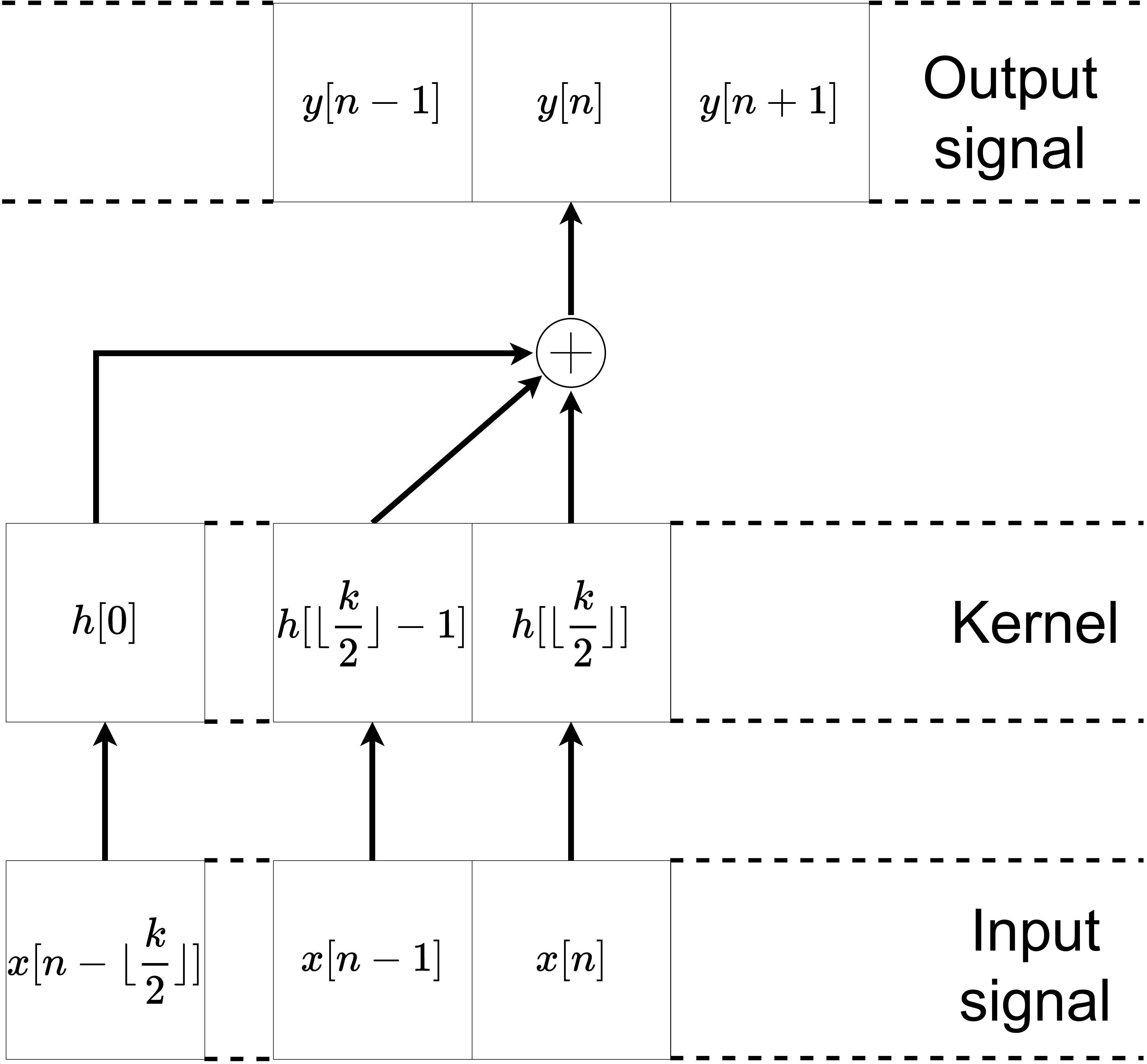}  
      \caption{Causal convolution.}
      \label{fig:causalconv} 
     \end{subfigure} \\
    \label{fig:convolutions}
\end{figure}

Our latest extension of the proposed model is to be able to run online and enable streamable extensions for real-time applications. To this end, we propose \csudoi which is a more shallow and efficient model which has the following differences against the improved model \sudoi.
\begin{itemize}
    \item We replace all non-causal convolutions with causal counterparts. Now the architecture does not depend on future samples in order to produce the estimated signal up to the current time-frame. A depiction of those two convolutional modules is presented in Figures \ref{fig:noncausalconv} and \ref{fig:causalconv} for the non-causal and the causal convolutional layers, respectively.
    \item In order to simplify the implementation and make those models more efficient in terms of memory footprint and execution time, we also remove all the normalization layers.
\end{itemize}

\section{Experimental Setup}
\label{sec:exp_setup}
\subsection{Audio source separation tasks}
\label{sec:exp_setup:datasets}
\noindent\textbf{Speech separation (2 active speakers):} We perform speech separation experiments using the publicly available WSJ0-2mix dataset \cite{hershey2016deepclustering} by following a similar setup with other studies \cite{luo2019dual,zeghidour2020wavesplit,liu2019DeepCASA}. Speaker mixtures are generated by randomly mixing speech utterances with two active speakers at random signal to noise ratios (SNR)s between $-5$ and $5$dB from the Wall Street Journal (WSJ0) corpus \cite{WSJ0}. \\

\noindent\textbf{Non-speech sound separation (2 active sources):} For our non-speech sound separation experiments we follow the exact same setup as in \cite{tzinis2019two} and utilize audio clips from the environmental sound classification (ESC50) data collection \cite{esc50} which consists of a wide variety of sounds (non-speech human sounds, animal sounds, natural soundscapes, interior sounds and. urban noises). For each data sample, two audio sources are mixed with a random SNR between $-2.5$ and $2.5$dB where each source belongs to a distinct sound category from a total of $50$. \\

\noindent\textbf{Universal sound separation (variable number of sources 1-4):} We also evaluate our models under a purely universal sound separation setup where multiple sound classes might be present and also we do not know how many sources are active in each input mixture. To that end, we use the FUSS benchmark dataset presented in \cite{wisdom2020FUSS}. FUSS contains sound clips that might contain at least one and up to four active sources per input mixture. Moreover, the sound clips represent a wide variety of real-world sounds including (speech, engine sounds, music, wind, rain, and many others). Also, the SNR distribution of the input sound mixtures is more realistic by capturing a wide range approximately from $-40$dB to $40$dB.

\subsection{Data pre-processing and generation}
We follow the same data augmentation process which was firstly introduced in \cite{tzinis2019two} and it has been shown beneficial in other recent studies \cite{zeghidour2020wavesplit}. We also normalize all processed audio clips by subtracting their mean and divide with their standard deviation.

\subsubsection{Fixed number of sources}
The process for generating a training mixture is the following: A) random choosing two sound classes (for non-speech sound separation) or speakers (for speech separation) B) random cropping of $4$sec segments from two sources audio files C) mixing the source segments with a random SNR (as specified in Section \ref{sec:exp_setup:datasets}). For each epoch, $20,000$ new training mixtures are generated. Validation and test sets are generated once with each one containing $3,000$ mixtures. We also downsample each audio clip to $8$kHz.

\subsubsection{Variable number of sources}
In order to be consistent with the state-of-the-art results on FUSS, we use the same dataset splits as the ones provided in \cite{wisdom2020FUSS}. The augmentation pipeline for each training mixture includes mixing sources from different training samples by sampling them uniformly over the batch. For each epoch, $20,000$ new training mixtures are generated. Validation and test sets contain $5,000$ and $3,000$ mixtures, respectively. Moreover, we also train and test keeping the same length of $10$secs at all clips as well as their sampling frequency which is $16$kHz.

\subsection{Training details}
\label{sec:exp_setup:train}

\subsubsection{Fixed number of sources}
All models are trained for $120$ epochs using a batch size equal to $4$. As a loss function we use the negative permutation-invariant \cite{Yu2017PIT} scale-invariant signal to distortion ratio (SI-SDR) \cite{le2019sdr}. The total loss for $N$ sources is computed as the average loss across each source as follows: For the $i$th source we define the loss between the clean signal $\textbf{s}$ and the estimates $\widehat{\textbf{s}}$ as: 
\begin{equation}
\label{eq:SISDR}
    \begin{gathered}
    \mathcal{L} = - \frac{1}{N} \sum_{i=1}^N \text{SI-SDR}(\textbf{s}_i^*, \widehat{\textbf{s}}_i) = - \frac{1}{N} \sum_{i=1}^N 10 \log_{10} \left( \frac{\| \alpha_i \textbf{s}_i^*\|^2}{\| \alpha \textbf{s}_i^* - \widehat{\textbf{s}}_i\|^2} \right),
    \end{gathered}
\end{equation}
where $\textbf{s}^*$ denotes the permutation of the sources that maximizes SI-SDR and $\alpha_i =  \widehat{\textbf{s}}_i^\top  \textbf{s}_i^* /\|\textbf{s}_i\|^2$ is a scalar used for making the loss invariant to the scale of the $i$th estimated source $\widehat{\textbf{s}}_i$.  During training, we use the Adam optimizer \cite{adam} with an initial learning rate set to $0.001$ and we decrease it by a 
factor of $5$ every $50$ epochs. By training the model for more epochs, using the same learning rate scheduler, does not yield any significant gains on the validation set. 

\subsection{Variable number of sources}
All models are trained for $40$ epochs using a batch size equal to $4$. In this case we assume that we know the maximum number of sources in each input mixture $\mathbf{x}$, e.g. $N$, but in reality the mixture might be comprised of only $N' < N$ active sources and $N - N'$ inactive sources. We follow a similar training procedure with the one presented in \cite{wisdom2020FUSS} and we force the network to produce zero outputs for the inactive slots after inferring the permutation that maximizes the total SNR. The loss for the active sources and inactive sources is defined in a permutation invariant sense as follows:
\begin{equation}
\label{eq:SNRmanysources}
    \begin{aligned}
    \mathcal{L} = \underset{\pi \in \Pi}{\min} \Bigg[ & - \frac{1}{N'} \sum_{i=1}^{N'} 10 \log_{10} \left( \frac{\| \textbf{s}_i\|^2 + \epsilon}{\| \textbf{s}_i - \widehat{\textbf{s}}_i^{(\pi)}\|^2 + \epsilon}  \right) + \\ &
      \frac{1}{N - N'} \sum_{i=N'+1}^{N} 10 \log_{10} \left( \|\widehat{\textbf{s}}_i^{(\pi)}\|^2 + \tau \| \mathbf{x}\|^2 + \epsilon \right) \Bigg],
    \end{aligned}
\end{equation}
where $\Pi$ is the set of all possible permutations of the estimated sources $\widehat{\mathbf{s}}$ and we assume that the first $N'$ target signals represent the active sources. The first part of the loss function forces the model to maximize the reconstruction fidelity of the $N'$ active sources while the second part forces it to produce close to zero energy estimates for the last $N-N'$, assuming that the best permutation is already in place. The constants $\epsilon = 10^{-9}$ and $\tau = 10^{-3}$ solve numerical stability issues created by zero target signals $\s$. During training, we use the Adam optimizer \cite{adam} with an initial learning rate set to $0.001$ and we decrease it by a 
factor of $2$ every $10$ epochs. Using the aforementioned scheduler we are able to obtain good performance with only $40$ epochs.

\subsection{Evaluation details}
\label{sec:exp_setup:eval}
In order to evaluate the performance of our models we use a stable version of the permutation invariant SI-SDR improvement (SI-SDRi) as proposed in \cite{wisdom2020FUSS}. The SI-SDRi metric has also been chosen in several studies in the source separation literature \cite{tzinis2019improving,tzinis2019two,luo2019dual,zeghidour2020wavesplit,liu2019DeepCASA,wisdom2020FUSS}. There are several other SNR-based metrics which also reflect the fidelity of the reconstructed sources such as signal to distortion ratio (SDR), signal to interference ratio (SIR) and signal to artifacts ratio (SAR) \cite{vincent2006performance}. SDR incorporates the errors from both artifacts and interference but is an overly optimistic performance measure compared to SI-SDR \cite{le2019sdr}. The improvement is defined as the gain that we get on the SI-SDR measure using the estimated signal instead of the mixture signal $\mathbf{x}$, as shown next:
\begin{equation}
\label{eq:SIDReval}
    \begin{aligned}
    \text{SI-SDRi} \left( \widehat{\textbf{s}}, \textbf{s}\right) = & \underset{\pi \in \Pi}{\max} \left[ \frac{1}{N'} \sum_{i=1}^{N'} 10 \log_{10} \left( \frac{\| \alpha_i \textbf{s}_i\|^2 + \epsilon}{\| \alpha_i \textbf{s}_i - \widehat{\textbf{s}}_i^{(\pi)}\|^2 + \epsilon}  \right) \right] \\
    & - \frac{1}{N'} \sum_{i=1}^{N'} 10 \log_{10} \left( \frac{\| \alpha_i \textbf{s}_i\|^2 + \epsilon}{\| \alpha_i \textbf{s}_i - \mathbf{x}\|^2 + \epsilon}  \right),
    \end{aligned}
\end{equation}
where $\alpha_i =    \textbf{s}_i^\top \widehat{\textbf{s}}_i^{(\pi*)} /\|\textbf{s}_i\|^2$, $\epsilon = 10^{-9}$ and $\pi*$ is the permutation that maximizes the average SI-SDRi over the active sources. For the case where non-active sources exist, thus, $N' < N$, we omit to compute the aforementioned metric as it provides infinity values. However, we follow a stricter evaluation metric than the one proposed in \cite{wisdom2020FUSS} in the sense that we do not exclude pairs of estimates-targets with low energy estimated sources, as we believe that the evaluation metric should also reflect a penalty for the cases where the model under-separates the input mixture, leading to $M < N'$ non-zero estimated sources. For the variable number of sources case and specifically for the single-source mixtures we simply report the maximum absolute SI-SDR obtained by each one of the estimated sources.

\subsection{\sudo configurations}
\label{sec:exp_setup:our_model_config}
We describe the default values for all proposed architectures \sudodot, \sudodoti and, \csudodoti. In the following experimental sections, all those values are going to be described as such, unless otherwise specified. For the encoder $\mathcal{E}$ and decoder modules $\mathcal{D}$ we use a kernel size $\KE=21$ for input mixtures sampled at $8$kHz and $\KE=41$ for $16$kHz. Also, the number of basis is equal to $\CE = 512$. For the configuration of each U-ConvBlock we set the input number of channels equal to $\Cout = 128$, the number of successive resampling operations equal to $Q = 4$ and, the expanded number of channels equal to $\Cin = 512$. In each subsampling operation we reduce the temporal dimension by a factor of $2$ and all depth-wise separable convolutions have a kernel length of $\Kin=5$ and a stride of $\Sin=2$. Only for the causal variation \csudodoti, we increase the number of input channels to $\Cout = 256$ and the default kernel length to $\Kin=11$ in order to increase the receptive field in shallower and more efficient architectures needed for real-time applications. For simplicity we use the following naming convention based on the number $B$ of U-ConvBlocks inside the separator module $\mathcal{S}$. Namely, \sudoxl, \sudol, \sudom, \sudos consist of $32$, $16$, $8$ and $4$ blocks, respectively. The same applies to the improved version \sudoi and its causal variation \csudodoti.  

\subsection{Literature models configurations}
\label{sec:exp_setup:literature}
We compare against the best configurations of some of the latest state-of-the-art approaches for speech \cite{luo2019convTasNet,luo2019dual}, universal \cite{tzinis2019two} and music \cite{defossez2019demucs} source separation. For a fair comparison with the aforementioned models we use the authors' original code, the best performing configurations for the proposed models as well as the suggested training process. For Demucs \cite{defossez2019demucs}, $80$ channels are used instead of $100$ in order to be able to train it on a single graphical processing unit (GPU). For the universal sound separation experiments with a variable number of sources, we compare against the reported numbers in \cite{wisdom2020FUSS}, where an enhanced variation of the ConvTasNet is used, namely, TDCN++.

\subsection{Measuring computational resources}
\label{sec:exp_setup:computational_resources}
One of the main goals of this study is to propose models for audio source separation which could be trained using limited computational resources and deployed easily on a mobile or edge-computing device  \cite{lane2016deepx_deeplearning_onmobiledevs}. Specifically, we consider the following aspects which might cause a computational bottleneck during inference or training:
\begin{enumerate}
    \item Number of executed floating point operations (FLOPs).
     \item Number of trainable parameters.
    \item Memory allocation required on the device for a single pass.
    \item Time for completing each process. 
\end{enumerate}
We are using various sampling profilers in Python using Pytorch \cite{pytorch} (version 1.7.1.) for tracing all the requirements of the non-causal models on a server with an Intel(R) Core(TM) i7-3820 @ 3.60GHz CPU and a GeForce GTX TITAN X GPU. For the causal models, we focus on the computational requirements on a much more resource-constrained hardware in order to show the applicability of our models for real-time source separation on devices used by typical users. Thus, we evaluate our causal models on a laptop with an Intel(R) Core(TM) i7-8750H @ 2.20GHz CPU. We measure the inference pass as a simple forward pass always on CPU while we consider that a backward pass is comprised of a forward pass in order to compute the estimated signals and the full back-propagation of the gradient and we measure its computational requirements on GPU.

\section{Results \& Discussion}
\label{sec:results}
In Tables \ref{tab:final_results_inference}, and \ref{tab:final_results_backwards} we show the separation performance under the speech and non-speech separation tasks for some of the most recent state-of-the-art models in the literature and the proposed \sudo configurations alongside computational requirements. Specifically, in Table \ref{tab:final_results_inference}, we focus on the computational aspects required during a forwards pass of those models on a CPU while in Table \ref{tab:final_results_backwards}, the same computational resource requirements are shown for a backward pass on GPU as well as the number of trainable parameters. It is easy to see that the proposed models can match and even outperform the separation performance of other several state-of-the-art models using orders of magnitude less computational requirements across the board.

\begin{figure}[!t]
    \centering
      \includegraphics[width=\linewidth]{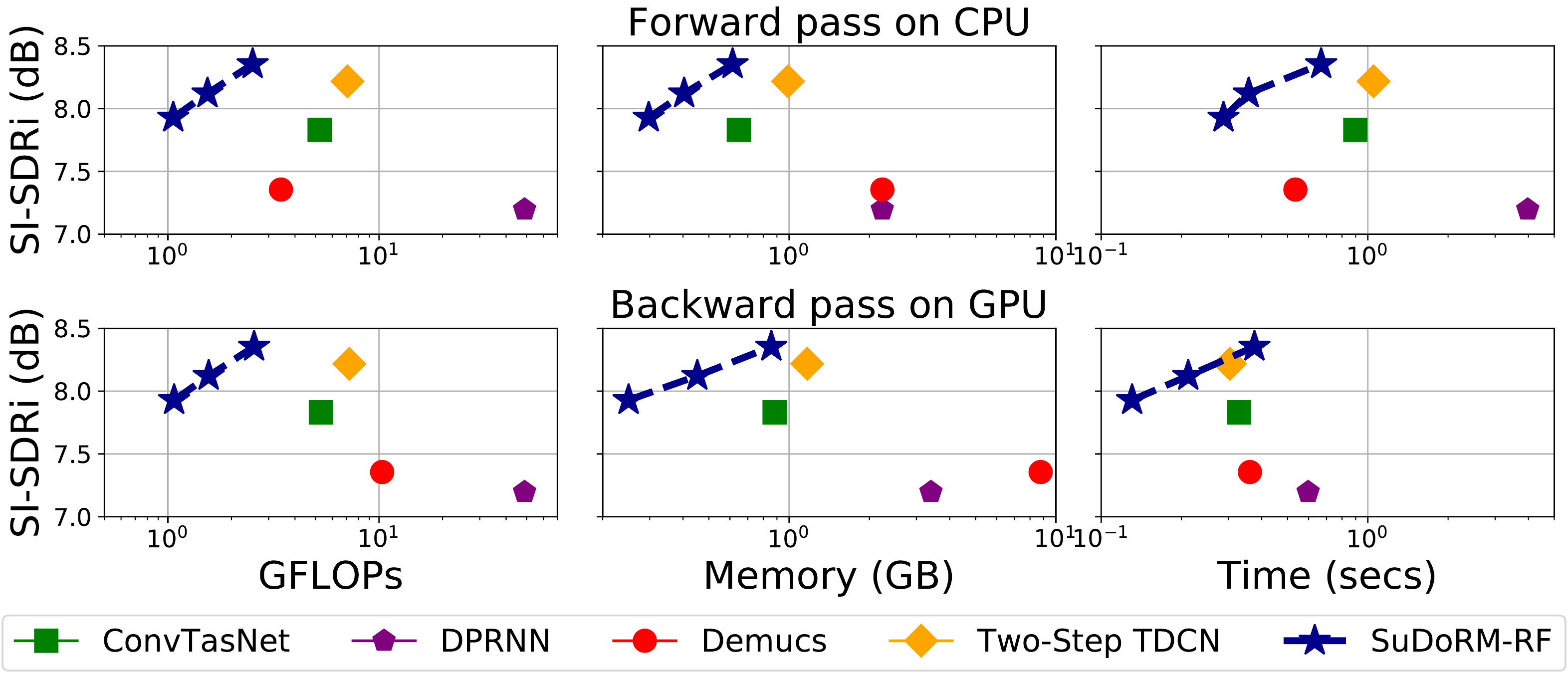}
      \caption{SI-SDRi non-speech sound separation performance on ESC50 vs computational resources with an input audio recording of $8000$ samples for all models. (Top row) computational requirements for a single forward pass on CPU (Bottom) for a backward pass on GPU. All x-axis are shown in log-scale while the $3$ connected blue stars correspond to the three \sudo configurations from Section \ref{sec:exp_setup:our_model_config}.}
      \label{fig:pareto}
\end{figure}

In Sections \ref{sec:results:FLOP}, \ref{sec:results:efficient_training}, \ref{sec:results:mem_requirements}, we focus on specific computational aspects for all the models presented in this study. Moreover, a better visualization for understanding the Pareto efficiency of the proposed architectures is displayed in Figure \ref{fig:pareto}. Specifically, we show for each model, its performance on non-speech sound separation vs a specific computational requirement. In Section \ref{sec:results:ablation}, we conduct a small ablation study to shed light on the most important aspects of the proposed \sudo models and how the performance is affected by changing the corresponding hyperparameters. Experiments for universal sound separation with a variable number of sources are presented in Section \ref{sec:results:fuss}. Finally, we present the causal setup of our model and evaluate it for different configurations under a two-speaker separation task in \ref{sec:results:causal_setup}.   

\begin{table*}[!t]
    \centering
    \begin{tabular}{l|c|c|c|c|c}
\toprule
\multirow{2}{*}{Model} & \multicolumn{2}{c|}{SI-SDRi (dB)} & \multirow{2}{*}{GFLOPs} & \multicolumn{1}{c|}{Mem.} & \multicolumn{1}{c}{Time}  \\
 & Speech & Non-speech & & \multicolumn{1}{c|}{(GB)} & \multicolumn{1}{c}{(sec)}  \\
\hlinewd{1pt}
ConvTasNet \cite{luo2019convTasNet} & 15.3* & 7.7 & 5.16 & 0.61 & 0.15   \\
Demucs \cite{defossez2019demucs} & 12.1 & 7.2 &  3.42 &  1.95  & 0.58   \\
DPRNN \cite{luo2019dual} & \textbf{18.8}* & 7.2 & 48.81  & 2.27 & 2.18  \\
Two-Step TDCN \cite{tzinis2019two} & 16.1* & 8.2 & 7.11 & 1.03 & 0.24 \\
\hlinewd{1pt}
\sudol & 17.0 & 8.4 & 2.45 & 0.79 & 0.17   \\
\sudom & 15.4 & 8.1 & 1.51 & 0.49 & 0.13   \\
\sudos & 13.4 & 7.9 & 1.04 & \textbf{0.30} & 0.09  \\
\hlinewd{1pt}
\sudoil & 17.0 & \textbf{8.6} & 2.11 & 0.79 & 0.17 \\
\hlinewd{1pt}
\sudoil GC & 12.5 & 8.5 & \textbf{0.69} & 0.87 & 0.20 \\
\hlinewd{1pt}
\csudoim & 10.6 & 4.6 & 2.14 & 0.39 & 0.08   \\
\csudois & 9.6 & 4.5 & 1.25 & \textbf{0.30} & \textbf{0.05}  \\
\bottomrule
\end{tabular}
\caption{SI-SDRi separation performance for the proposed models and models in the literature on both separation tasks (speech and non-speech) alongside their computational requirements for performing inference on an Intel(R) Core(TM) i7-3820 @ 3.60GHz CPU for one second of input audio or equivalently $8000$ samples. * We assign the maximum SI-SDRi performance obtained by our runs and the reported number on the corresponding paper.}
\label{tab:final_results_inference}
\end{table*}

\begin{table*}[!t]
    \centering
    \begin{tabular}{l|c|c|c|c|c|c}
\toprule
\multirow{2}{*}{Model} & \multicolumn{2}{c|}{SI-SDRi (dB)} &  Params & \multirow{2}{*}{GFLOPs} & \multicolumn{1}{c|}{Mem.} & \multicolumn{1}{c}{Time}  \\
 & Speech & Non-sp. & (1e6) & & \multicolumn{1}{c|}{(GB)} & \multicolumn{1}{c}{(sec)}  \\
\hlinewd{1pt}
ConvTasNet \cite{luo2019convTasNet} & 15.3* & 7.7 & 5.05 & 21.7 &  0.68 &  0.14  \\
Demucs \cite{defossez2019demucs} & 12.1 & 7.2 & 415.09 & 17.65 & 8.77 & 0.24  \\
DPRNN \cite{luo2019dual} & \textbf{18.8}* & 7.2 & 2.63 & 135.88 & 3.28 & 0.32  \\
Two-Step TDCN \cite{tzinis2019two} & 16.1* & 8.2 & 8.63 & 28.36& 1.17& 0.22 \\
\hlinewd{1pt}
\sudol & 17.0 & 8.4 & 2.72 & 17.19 & 0.99 & 0.27  \\
\sudom & 15.4 & 8.1 & 1.42 & 9.47 & 0.33 &0.11  \\
\sudos & 13.4 & 7.9 & 0.79 & 5.62 & 0.28 & 0.08  \\
\hlinewd{1pt}
\sudoil & 17.0 & \textbf{8.6} & 2.72 & 16.23 & 0.99 & 0.27  \\
\hlinewd{1pt}
\sudoil GC & 12.5 & 8.5 & \textbf{0.30} & \textbf{2.72} & 1.21 & 0.35  \\
\hlinewd{1pt}
\csudoim & 10.6 & 4.6 & 2.81 & 11.43 & 0.29 & 0.09  \\
\csudois & 9.6 & 4.5 & 1.63 & 6.27 & \textbf{0.17} & \textbf{0.05}  \\
\bottomrule
\end{tabular}
\caption{SI-SDRi separation performance for the proposed models and models in the literature on both separation tasks (speech and non-speech) alongside their computational requirements for performing a backward pass on a GeForce GTX
TITAN X GPU for one second of input audio or equivalently $8000$ samples. * We assign the maximum SI-SDRi performance obtained by our runs and the reported number on the corresponding paper.}
\label{tab:final_results_backwards}
\end{table*}

\subsection{Floating point operations (FLOPs)}
\label{sec:results:FLOP}
Different devices (CPU, GPU, mobiles, etc.) have certain limitations on their FLOPs throughput capacity. In the case of an edge device, the computational resource one might be interested in is the number of FLOPs required during inference. On the other hand, training on cloud machines might be costly if a huge number of FLOPs is needed in order to achieve high separation performance. As a result, it is extremely important to be able to train and deploy models which require a low number of computations \cite{howard2017mobilenets}. We see from the first column of Figure \ref{fig:pareto} that \sudo models scale well as we increase the number of U-ConvBlocks $B$ from $4\rightarrow8\rightarrow16$. Furthermore, we see from Tables \ref{tab:final_results_inference} and \ref{tab:final_results_backwards} that for both forward and backward passes, correspondingly, the family of the proposed \sudo models appear more Pareto efficient in terms of SI-SDRi performance vs Giga-FLOPs (GFLOPs) and time required compared to the other state-of-the-art models which we take into account. Specifically, the DPRNN model \cite{luo2019dual} which performs sequential matrix multiplications (even with a low number of parameters) requires at least $45$ times more FLOPs for a single pass compared to \sudos while performing worse when trained for the same number of epochs under the non-speech separation task.

Moreover, we see from both Tables \ref{tab:final_results_inference} and \ref{tab:final_results_backwards} that the improved version \sudoi achieves similar or even better results than the original version of the proposed model \sudo with a lower number of FLOPs both in forward and backward for a similar number of parameters and execution time. A significant drop in the absolute number of FLOPs is also obtained by combining the group communication mechanism proposed in \cite{luo2020groupcomm} with \sudoi. However, that does not automatically entail a lower execution time. The causal variations \csudoi perform competitively with the same number of FLOPs but they are still performing much worse than all the other non-causal models showing that there is much room for improvement.

\subsubsection{Cost-efficient training}
\label{sec:results:efficient_training}
\begin{figure}[!htb]
    \centering
      \includegraphics[width=\linewidth]{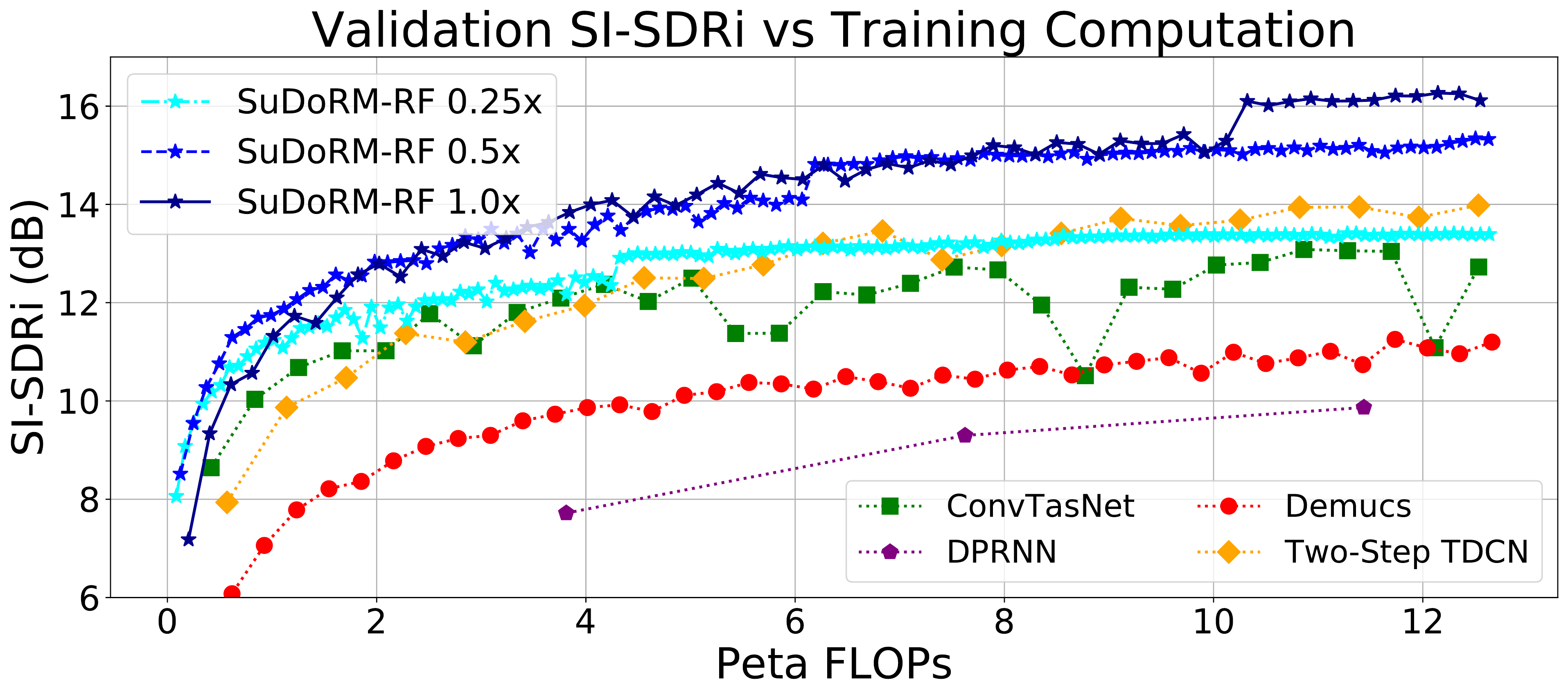}
      \caption{Validation SI-SDRi separation performance for speech-separation vs the number of FLOPs executed during training. All models are trained using batches of $4$ mixtures with $32,000$ time-samples each. Each point corresponds to a completed training epoch.}
      \label{fig:costefficienttraining}
    \vspace{-15pt}
\end{figure}

Usually one of the most detrimental factors for training deep learning models is the requirement of allocating multiple GPU devices for several days or weeks until an adequate performance is obtained on the validation set. This huge power consumption could lead to huge cloud services rental costs and carbon dioxide emissions \cite{cai2019onceandforall}. In Figure \ref{fig:costefficienttraining}, we show the validation SI-SDRi performance for the speech separation task which is obtained by each model versus the total amount of FLOPs performed. For each training epoch, all models perform updates while iterating over $20,000$ audio mixtures. Notably, even the original \sudo models outperform all other models in terms of cost-efficient training as they obtain better separation performance while requiring significantly fewer training FLOPs. For instance, \sudol obtains $\approx 16$dB in terms of SI-SDRi compared to $\approx 10$dB of DPRNN \cite{luo2019dual} which manages to complete only $3$ epochs given the same number of training FLOPs.   

\subsection{Trainable parameters}
\label{sec:results:trainable_parameters}
From Table \ref{tab:final_results_backwards} it is easy to see that \sudo architectures are using orders of magnitude fewer parameters compared to the U-net architectures like Demucs \cite{defossez2019demucs} where each temporal downsampling is followed by a proportional increase to the number of channels. Moreover, the upsampling procedure inside each U-ConvBlock does not require any additional parameters. The \sudo models seem to increase their effective receptive field with significantly fewer parameters compared to dilated convolutional architectures \cite{luo2019convTasNet,tzinis2019two}. Notably, our largest model \sudol matches the relatively low number of parameters of the DPRNN \cite{luo2019dual} model which is based on stacked RNN layers. Group communication combined with \sudoi is one of the most effective ways to reduce the number of trainable parameters caused by the bottleneck dense layers between the U-ConvBlocks. Essentially, the \sudoil GC model with $B=16$ number of U-ConvBlocks has less than half of the parameters of a shallow original \sudos model with only $B=4$ U-ConvBlocks.

\subsection{Memory requirements}
\label{sec:results:mem_requirements}
In most of the studies where efficient architectures are introduced \cite{howard2017mobilenets,chollet2017xception_depthwiseseparable,mehta2019espnetv2,yu2019slimmablenets} authors are mainly concerned with the total number of trainable parameters of the network. The same applies to efficient architectures for source separation \cite{luo2019convTasNet,luo2019dual,maldonado2020lightweight}. However, the trainable parameters comprise only a small portion of the total amount of memory required for a single forward or backward pass. The space complexity could easily be dominated by the storage of intermediate representations and not the actual memory footprint of the model itself. The latter could become even worse when multiple skip connections are present, gradients from multiple layers have to be stored or implementations require augmented matrices (dilated, transposed convolutions, etc.).

From Tables \ref{tab:final_results_inference} and \ref{tab:final_results_backwards}, we see that \sudoi and the initial \sudo models have almost the same memory footprint. However, when combining the \sudoi with the group communication (GC) mechanism we see that even if the number of trainable parameters is significantly reduced then still the actual memory requirement increases which positively validates our hypothesis that the actual memory requirement in GBs is a very important metric to show when proposing new efficient models. The causal variation of our models \csudoi is the more light-weight model that we propose which also secures a very competitive inference time. 

In Figure \ref{fig:pareto}, we see that even the original \sudo models are more Pareto efficient in terms of the memory required compared to the dilated convolutional architectures of ConvTasNet \cite{luo2019convTasNet} and Two-Step TDCN \cite{tzinis2019two} where they require an increased network depth in order to increase their receptive field. Although \sudo models do not perform downsampling in every feature extraction step as Demucs \cite{defossez2019demucs} does, we see that the proposed models require orders of magnitude less memory especially during a backward update step as the number of parameters in Demucs is significantly higher. Finally, \sudo models have a smaller memory footprint because the encoder $\mathcal{E}$ performs temporal downsampling by a factor of $\operatorname{div}\lp \KE, 2\rp=10$ compared to DPRNN \cite{luo2019dual} which does not reduce the temporal resolution at all.

\subsection{Ablation study on WSJ0-2mix}
\label{sec:results:ablation}
We perform an ablation study in order to show how different parameter choices in \sudoi models affect the separation performance. In order to be directly comparable with the numbers reported by several other studies  \cite{luo2019convTasNet,luo2019dual,zeghidour2020wavesplit,liu2019DeepCASA}, we train our models for $200$ epochs and test them using the given data splits from WSJ0-2mix dataset \cite{hershey2016deepclustering}. The results are shown in Table \ref{tab:ablation_study}.

We see that a significant aspect is the stride of the encoder and decoder which is always defined as $\KE // 2$. By decreasing the size of the stride we force the model to perform more computations and also estimate the signal in a more fine-grained scale closer to the time-domain resolution leading to better results which is also consistent with other studies \cite{kavalerov2019universal}. Moreover, we see that the GLN significantly helps our model to reach a better solution compared to the simple layer norm, presumably acting as a better regularizer in between the intermediate activations. Furthermore, when keeping all the other parameters the same except for the number of U-ConvBlocks $B$ and the number of resampling procedures $Q$ we see one of the most important aspects of the \sudoi model which is the benefit in the receptive field that we are getting by analyzing the signal at multiple scales. Specifically, we see that a model with $B=18$ and $Q=7$ outperforms a deeper model in terms of U-ConvBlocks $B=20$ which only processes the signal at $Q=2$ more scales. We need to underline that increasing the parameter $Q$ does not lead to significant computational requirements mainly because of the downsampling operation which is relatively a cheap way to increase the receptive field of the model without carrying its whole information end-to-end.

\begin{table}[!t]
    \centering
    \begin{tabular}{c|c|c|c|c|c|c}
\toprule
$\KE$ & $\Cout$ & $B$ & $Q$ & Normalization  & SI-SDRi \\
\hlinewd{1pt}
17 & 128 & 16 & 4& LN   & 15.9 \\
\hline
17 & 128 & 16 & 4& GLN & 16.8 \\
\hline
21 & 256 & 20 & 4& GLN   & 17.7 \\
\hline
41 & 256 & 32 & 4 & GLN  & 17.1 \\
\hline
41 & 256 & 20 & 4 & GLN & 16.8 \\
\hline
21 & 512 & 18 & 7 & GLN & 18.0 \\
\hline
21 & 512 & 20 & 2 & GLN & 17.4 \\
\hline
21 & 512 & 34 & 4 & GLN & 18.9 \\
\bottomrule
\end{tabular}
\caption{SI-SDRi separation performance on WSJ0-2mix for various parameter configurations of \sudoi models. GLN corresponds to the global layer normalization as described in Equation \ref{eq:gln} and LN corresponds to the classic layer normalization layer proposed in \cite{ba2016layernormalization} and explained in Equation \ref{eq:ln}. All the other parameters have the same values as described in Section \ref{sec:exp_setup:our_model_config}.}
\label{tab:ablation_study}
\end{table}

\subsection{Variable number of sources}
\label{sec:results:fuss}
In Table \ref{tab:variable_number_of_sources} we report the performance of the proposed \sudo models under a universal sound separation task with a varying number of sources in each mixture. We always assume that the maximum number of active sources in each mixture is $N=4$ and we measure the performance on the same dataset splits where mixtures with $N' \in \{1, 2 , 3, 4\}$ active sources. We see that by increasing our \sudo and \sudoi model sizes to match the size of TDCNN++ \cite{wisdom2020FUSS}, we can match its performance which is also the current state-of-the-art performance in universal sound separation with a variable number of sources. For the single source mixtures we see that our models perform worse than the TDCNN++, however, above 25dB it is really difficult even for a human being to understand the nuance artifacts which are barely audible. Our \sudoixl outperforms the state-of-the-art for the most difficult case where $N'=4$ sources are active. Moreover, by using the \sudoil GC, we match the performance obtained by the state-of-the-art with a significantly fewer parameters. We also mention that our models perform worse than the state-of-the-art for the cases where $N'=2$ and $N'=3$ because we also penalize the performance of our models in the cases of under-separation which has been reported as $25\%$ in \cite{wisdom2020FUSS}. Namely, for the pairs of corresponding estimates-targets with low-energy estimates, the state-of-the-art numbers only consider the pairs with estimated sources which have power higher than $20$ dB above the power of the quietest non-zero reference source. However, we always include all pairs of estimates-targets if a target source is non-zero as we believe that this is a more appropriate metric to use. We also like to note that we did not notice any significant performance improvement when we use mixture consistency \cite{wisdom2019differentiable} at the output estimates of our models where the separated sources are forced to sum up to the input mixture using a projection layer.

\begin{table}[!t]
    \centering
\begin{tabular}{l|c|ccc|c}
\toprule
\multirow{2}{*}{Model} & SI-SDR (dB) & \multicolumn{4}{c}{SI-SDRi (dB)}  \\
 &  $N'=1$ & $N'=2$ & $N'=3$ & $N'=4$ & Avg. (2-4)\\
\hlinewd{1pt}
TDCN++ \cite{wisdom2020FUSS} & \textbf{35.5} & \textbf{11.2}* & \textbf{11.6}* & 7.4  & \textbf{10.1} \\
\hlinewd{1pt}
SuDoRM-RF 2.0x & 22.2 & 9.9 & 9.6 & 7.0 & 8.8 \\
SuDoRM-RF++ 2.0x & 25.9 & 10.9 & 10.6 & \textbf{7.8} & 9.8 \\
SuDoRM-RF++ 1.0x GC & 22.9 & 9.9 & 9.9 & 7.3 & 9.0 \\
SuDoRM-RF++ 0.5x & 20.0 & 8.4 & 8.3 & 6.0 & 7.6 \\
\bottomrule
\end{tabular}
\caption{SI-SDR and SI-SDRi separation performance on the FUSS dataset for a variable number of sources. Specifically, we report the absolute SI-SDR ($N'=1$) for the reconstruction of single-source mixtures. For mixtures with multiple active sources $N' > 1$ we measure the SI-SDRi performance of the reconstructed sources wrt the input mixture. *Metrics do not consider pairs of corresponding estimates-target sources with low energy estimates, where a $25\%$ of under-separation is reported in the cases of $N' \in \{2, 3, 4\}$.}
\label{tab:variable_number_of_sources}
\end{table}

\begin{table}[!t]
    \centering
    \begin{tabular}{c|c|c|c}
\toprule
$B$ & $\Kin$ & SI-SDRi & Time (ms) \\
\hlinewd{1pt}
\multirow{3}{*}{4} & 3 & 8.4 & 50.8 \\ 
 & 5 & 9.1 & 50.4 \\ 
 & 11 & 9.5 & 88.7 \\ 
\hline
\multirow{3}{*}{8} & 3 & 9.6 & 90.6 \\ 
 & 5 & 10.1 & 88.2 \\ 
 & 11 & 10.3 & 165.9 \\ 
\bottomrule
\end{tabular}
\caption{SI-SDRi separation performance on WSJ0-2mix for various parameter configurations of causal \csudoi models alongside their inference time for $8000$ input samples on a laptop with an Intel(R) Core(TM) i7-8750H @ 2.20GHz CPU. For reference, one of the most current state-of-the-art speech denoisers, namely, real-time Demucs \cite{defossez2020realtimedemucs} runs on the same setup at $92.3$ms.}
\label{tab:causal_results}
\end{table}

\subsection{Causal setup}
\label{sec:results:causal_setup}
In Table \ref{tab:causal_results} we report the performance of the proposed causal version of \sudo models under a separation setup where two speakers are active using the WSJ0-2mix dataset \cite{hershey2016deepclustering} alongside the inference time on a laptop for $1$ second of input audio sampled at $8$kHz. We see that we are able to obtain competitive separation performance for all configurations while remaining $\approx$ $10$ to $20$ times faster than real time.

\section{Conclusions}
\label{sec:conclusions}
In this study, we have introduced the \sudo network, a novel architecture for efficient universal sound source separation. Moreover, we have presented several improvements on the original model including \sudoi which directly estimates the of the latent representations of the models, a variation which shares parameters across sub-bands as well as \csudoi which is causal and enables real-time inference. The proposed model is based on the U-ConvBlock which is capable of extracting multi-resolution temporal features through successive depth-wise convolutional downsampling of intermediate representations and aggregates them using a non-parametric interpolation scheme. In this way, \sudo models are able to significantly reduce the required number of layers in order to effectively capture long-term temporal dependencies. We show that these models can perform similarly or even better than recent state-of-the-art models while requiring significantly less computational resources in FLOPs, memory and, time for experiments with a fixed and a variable number of sources.

\bibliographystyle{spmpsci}      
\bibliography{refs}

\end{document}